\documentclass[%
 reprint,
superscriptaddress,
nofootinbib,
 amsmath,amssymb,
 aps,
pra,
]{revtex4-2}

\usepackage{graphicx}
\usepackage{dcolumn}
\usepackage{bm}
\usepackage{hyperref}

\begin{document}

\title{QUICK$^3$ - Design of a satellite-based quantum light source for quantum communication and extended physical theory tests in space}

\author{Najme Ahmadi}
\thanks{These authors contributed equally.}
\affiliation{Institute of Applied Physics, Abbe Center of Photonics, Friedrich Schiller University Jena, 07745 Jena, Germany}
\author{Sven Schwertfeger}
\thanks{These authors contributed equally.}
\affiliation{Ferdinand-Braun-Institut (FBH), 12489 Berlin, Germany}
\author{Philipp Werner}
\affiliation{Institut f\"ur Luft- und Raumfahrt, Technische Universit\"at Berlin, 10587 Berlin, Germany}
\author{Lukas Wiese}
\affiliation{Institut f\"ur Luft- und Raumfahrt, Technische Universit\"at Berlin, 10587 Berlin, Germany}
\author{Joseph Lester}
\affiliation{Institut f\"ur Luft- und Raumfahrt, Technische Universit\"at Berlin, 10587 Berlin, Germany}
\author{Elisa Da Ros}
\affiliation{Ferdinand-Braun-Institut (FBH), 12489 Berlin, Germany}
\affiliation{Department of Physics, Humboldt University of Berlin, 12489 Berlin, Germany}
\author{Josefine Krause}
\affiliation{Institute of Applied Physics, Abbe Center of Photonics, Friedrich Schiller University Jena, 07745 Jena, Germany}
\affiliation{Max Planck School of Photonics, 07745 Jena, Germany}
\author{Sebastian Ritter}
\affiliation{Institute of Applied Physics, Abbe Center of Photonics, Friedrich Schiller University Jena, 07745 Jena, Germany}
\affiliation{Max Planck School of Photonics, 07745 Jena, Germany}
\author{Mostafa Abasifard}
\affiliation{Institute of Applied Physics, Abbe Center of Photonics, Friedrich Schiller University Jena, 07745 Jena, Germany}
\author{Chanaprom Cholsuk}
\affiliation{Institute of Applied Physics, Abbe Center of Photonics, Friedrich Schiller University Jena, 07745 Jena, Germany}
\author{Ria G. Kr\"amer}
\affiliation{Institute of Applied Physics, Abbe Center of Photonics, Friedrich Schiller University Jena, 07745 Jena, Germany}
\author{Simone Atzeni}
\affiliation{Istituto di Fotonica e Nanotecnologie (IFN), Consiglio Nazionale delle Ricerche (CNR), 20133 Milan, Italy}
\author{Mustafa G\"undo\u{g}an}
\affiliation{Ferdinand-Braun-Institut (FBH), 12489 Berlin, Germany}
\affiliation{Department of Physics, Humboldt University of Berlin, 12489 Berlin, Germany}
\author{Subash Sachidananda}
\affiliation{Centre for Quantum Technologies, Department of Physics, National University of Singapore, 117543 Singapore, Singapore}
\author{Daniel Pardo}
\affiliation{Centre for Quantum Technologies, Department of Physics, National University of Singapore, 117543 Singapore, Singapore}
\author{Stefan Nolte}
\affiliation{Institute of Applied Physics, Abbe Center of Photonics, Friedrich Schiller University Jena, 07745 Jena, Germany}
\affiliation{Fraunhofer-Institute for Applied Optics and Precision Engineering IOF, 07745 Jena, Germany}
\author{Alexander Lohrmann}
\affiliation{SpeQtral Pte.\ Ltd., 138632 Singapore, Singapore}
\author{Alexander Ling}
\affiliation{Centre for Quantum Technologies, Department of Physics, National University of Singapore, 117543 Singapore, Singapore}
\author{Julian Bartholom\"aus}
\affiliation{Institut f\"ur Luft- und Raumfahrt, Technische Universit\"at Berlin, 10587 Berlin, Germany}
\author{Giacomo Corrielli}
\affiliation{Istituto di Fotonica e Nanotecnologie (IFN), Consiglio Nazionale delle Ricerche (CNR), 20133 Milan, Italy}
\author{Markus Krutzik}
\affiliation{Ferdinand-Braun-Institut (FBH), 12489 Berlin, Germany}
\affiliation{Department of Physics, Humboldt University of Berlin, 12489 Berlin, Germany}
\author{Tobias Vogl}
\email{tobias.vogl@uni-jena.de}
\affiliation{Institute of Applied Physics, Abbe Center of Photonics, Friedrich Schiller University Jena, 07745 Jena, Germany}
\affiliation{Fraunhofer-Institute for Applied Optics and Precision Engineering IOF, 07745 Jena, Germany}

\date{\today}

\begin{abstract}
Modern quantum technologies have matured such that they can now be used in space applications, e.g., long-distance quantum communication. Here, we present the design of a compact true single photon source that can enhance the secure data rates in satellite-based quantum key distribution scenarios compared to conventional laser-based light sources. Our quantum light source is a fluorescent color center in hexagonal boron nitride. The emitter is off-resonantly excited by a diode laser and directly coupled to an integrated photonic processor that routes the photons to different experiments performed directly on-chip: (i) the characterization of the single photon source and (ii) testing a fundamental postulate of quantum mechanics, namely the relation of the probability density and the wave function (known as Born's rule). The described payload is currently being integrated into a 3U CubeSat and scheduled for launch in 2024 into low Earth orbit. We can therefore evaluate the feasibility of true single photon sources and reconfigurable photonic circuits in space. This provides a promising route toward a high-speed quantum network.
\end{abstract}


\maketitle


\section*{Introduction}
With the recent rapid advances, modern quantum technologies have started to transition into a phase of commercialization \cite{Ac_n_2018}. The first major impact of quantum technologies on our globalized and interconnected world could have quantum key distribution (QKD) with its ultimate level of secrecy and privacy. The security of QKD is based on physical laws and therefore independent of the available computational power any potential eavesdropper might have. For many QKD protocols, the information is encoded into single photon states of light. This information can be neither fully readout due to the Heisenberg uncertainty nor copied perfectly due to the no-cloning theorem \cite{RevModPhys.74.145}. Any attempts in this direction inevitably change the quantum state which reveals the eavesdropping.\\
\indent Crucial for QKD is the encoding in single photons to prevent photon-number-splitting (PNS) attacks \cite{L_tkenhaus_2002}, where only a subset of multi-photon pulses is intercepted which in principle does not alter the quantum state (and thereby also not reveal the presence of the eavesdropper). While there has been a lot of research on the development of single photon sources \cite{10.1038/nphoton.2016.186,10.1038/nnano.2017.218}, building an ideal single photon source remains a technical challenge.\\
\indent The transmission of the information carriers from the sender to the receiver is another critical aspect of QKD. Due to the finite absorption and scattering losses in silica fibers, the maximal communication distance is limited to a few hundred kilometers for standard QKD protocols \cite{PhysRevLett.121.190502}. More advanced protocols, such as twin-field QKD, were able to push this to 830 km \cite{Wang2022}, but the exponential damping of light in fibers will always be a fundamental limit. Intercontinental distances can be bridged with satellites \cite{10.1038/nature23655,PhysRevLett.120.030501}, as the scattering in the atmosphere above 10 km becomes negligible \cite{RevModPhys.94.035001}.\\
\indent A global quantum-secured internet is likely going to be hybrid, where metropolitan fiber networks are connected with satellite links. Such quantum network will also use space-born quantum memories \cite{PhysRevA.91.052325,Gundogan2021}, either to build quantum repeaters, or to store a photon on an orbiting satellite to distribute quantum information between ground stations that do not have a common line of sight to the satellite. The efficient coupling between different quantum systems (e.g., single photon sources and quantum memories) is therefore essential.\\
\indent The first generation of (already launched) QKD satellites utilize weak coherent laser pulses as their light source \cite{10.1038/nature23655,Takenaka2017}. This comes at the expense of a low mean photon number (to reduce the amount of multi-photon pulses). This naturally reduces the data rate, as empty pulses do not carry any information. The mean photon number can be increased with decoy protocols \cite{PhysRevLett.94.230504}, which use weak coherent states with varying intensities. The PNS attack alters the photon statistics at the receiver, which again reveals the eavesdropping attempt. Even with decoy states, however, the secure data rate will always be smaller compared to using ideal single photons. Given the short flyover times of a QKD satellite in low Earth orbit (LEO), which is typically below 5 minutes \cite{10.1038/nature23655}, high data rates are crucial. The next generation QKD satellites should therefore feature higher data rates to expand the possible application scenarios.\\
\indent In the QUICK$^3$ mission\footnote{QUICK$^3$ stands for \textit{\textbf{QU}antenphoton\textbf{I}s\textbf{C}he \textbf{K}omponenten f\"ur sichere \textbf{K}ommunikation mit \textbf{K}leinsatelliten}.}, we aim to demonstrate a universal quantum architecture on a satellite. The quantum light source is based on a color center in the 2D material hexagonal boron nitride (hBN) \cite{nnano.2015.242} directly coupled to an actively switchable photonic integrated circuit (PIC) \cite{Corrielli2021}. The single photon emitter is excited by an extended cavity diode laser (ECDL) at 698 nm. To verify that our photon source emits single photons in space, we utilize single photon detectors based on the heritage from the SpooQy-1 mission \cite{Villar:20}. While our component test provides an important milestone toward establishing long-distance QKD links with true single photon sources that can enhance the data rate, our mission also tests extended quantum theories \cite{PhysRevResearch.3.013296} in space. For the satellite bus we resort to using a commercial multi-purpose 3U CubeSat platform. Instead of a dedicated development of space-certified components, we take the new space approach, where we test our components under space environment conditions. While this approach bears a higher risk, it allows us to close the gap on the currently designed second generation QKD satellite missions. In the following, we present the mission design, similarly as it has been done for previous satellite and sounding rocket missions \cite{Schkolnik2017,Neumann2018,Kerstel2018} and show preliminary space-qualification results.
\begin{figure*}[t]
\includegraphics[width=0.8\textwidth]{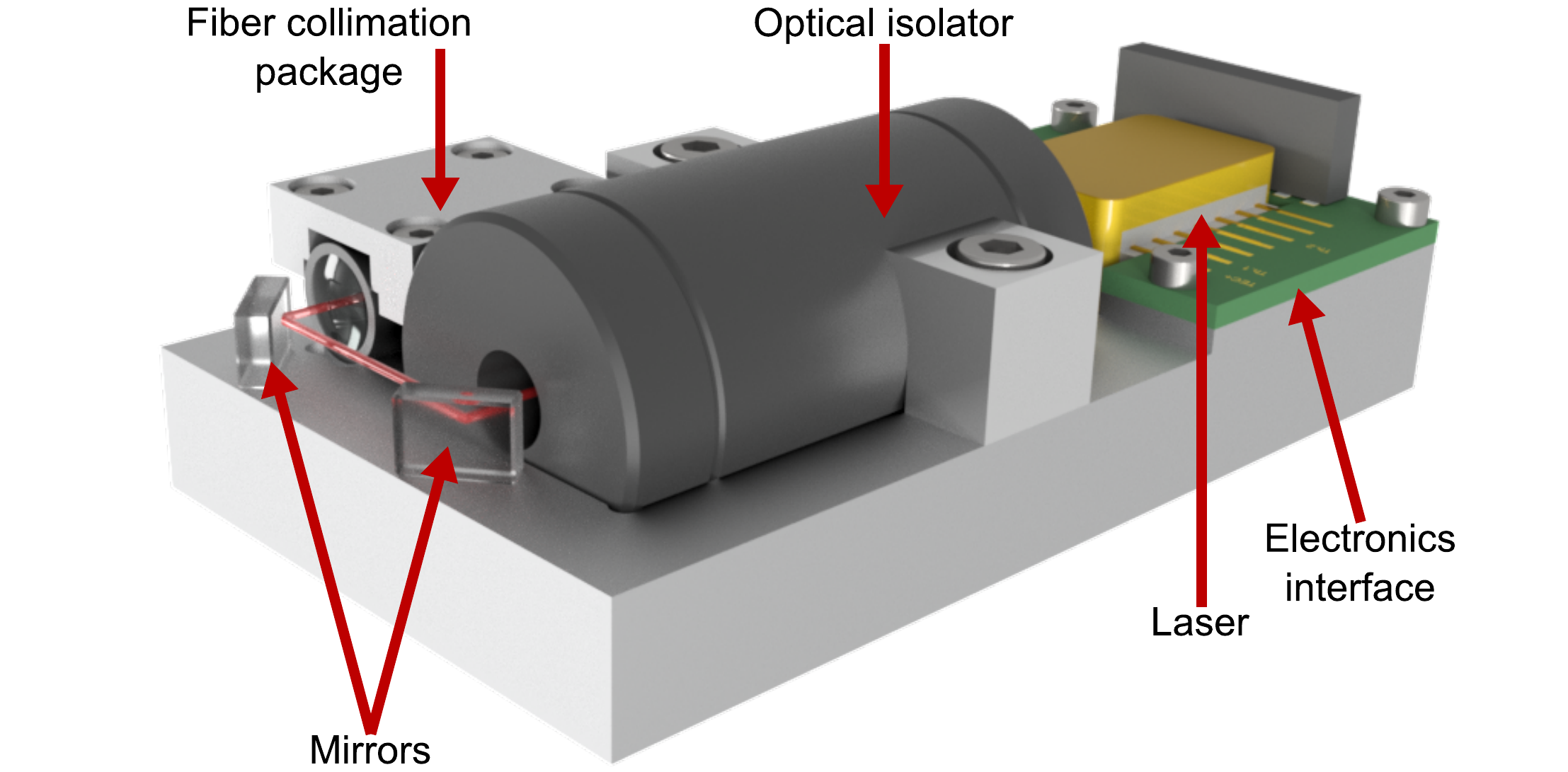}
\caption{Concept of the excitation laser system. The butterfly-packaged ECDL emits at 698 nm into the optical isolator that prevents optical feedback. With two mirrors, the beam is steered into a fiber collimation package, such that the fiber can guide the laser to the quantum photonics module.}
\label{Fig:1}
\end{figure*}

\section*{Mission overview}
The QUICK$^3$ mission aims at the demonstration of our components for future use in satellite-based quantum networks. The mission parameters were chosen to be comparable to that of a QKD scenario \cite{10.1038/nature23655,Takenaka2017}, i.e., in LEO with an altitude of 400-650 km. We require a minimal experimental time in orbit of at least one year. To study long-term effects (e.g., how long specific sub-systems survive), an orbital lifetime of at least three years is aspired. The re-entry into the atmosphere must be within 25 years to comply with space debris mitigation policies. This reduces the altitude window for our specific spacecraft to 487-604 km (see Supplementary Section S1).\\
\indent The orbital inclination should allow us to connect to the satellite from our ground stations located in Berlin, Germany (52.5$^\circ$ N) and Longyearbyen, Norway (78.2$^\circ$ N). This requirement already eliminates deployment from the International Space Station, which has been used for CubeSats launches in the past \cite{Villar:20,10.1117/1.JRS.13.032504}. To avoid short contact durations with the ground station in Berlin, the lower limit for the orbital inclination is 64$^\circ$. Supplementary Section S2 shows the ground trajectories for orbital inclinations of 53$^\circ$ and 64$^\circ$. With the latter orbit, we have eight (instead of six) passes per day with total contact times of 80 minutes (instead of 65 minutes). In addition, with 64$^\circ$ there are fewer passes with high elevation (near zenith), which makes it difficult for the UHF (ultra high frequency) antennas in our ground station to follow pointing (near zenith). As an exotic (non-standard) orbit is undesirable, this essentially limits the potential options to the widely used Sun-synchronous orbit (SSO). This is also a good choice for QKD scenarios, as the SSO is a near polar orbit and has therefore great geographical coverage.\\
\indent For the spacecraft, we have chosen the multipurpose 3U nano-satellite bus from NanoAvionics. The satellite features a payload volume of about 1.5U and an available payload mass of 2.3 kg. The satellite and all sub-systems are flight-proven (TRL 9). The six batteries with a capacity of 69 Wh are recharged by two deployable solar panels and regulated by the satellite's internal electrical power system. For our orbit, the solar panels are able to generate 15.2 W average orbital power. The data exchange is carried out via a UHF radio module (for telemetry data) and an S-band transceiver (for experimental data). The satellite also has an attitude determination and control system, which detumbles the spacecraft after deployment from the launch vehicle and can maintain a stable satellite orientation for efficient power generation and data transfer. The interface to our payload is managed through a dedicated payload controller.\\
\indent Our payload consists of the excitation laser system, the quantum emitter directly coupled to the photonic integrated circuit, a single photon detection system, and the corresponding driving electronics as well as a payload controller that serves as the interface to the satellite bus. The optical payload components are connected with each other via single mode fibers. The experiments that we want to carry out are (i) the in-orbit verification of the single photon source as a use-case for future satellite-based QKD, and (ii) testing a fundamental postulate of quantum physics, namely that the probability density of a quantum object is defined by the absolute square of its wave function (known as Born's rule), in space as a science-case. While Born's rule is fundamental to the framework of quantum mechanics, it cannot be derived from first principles without making other assumptions on the mathematical structure of the measurement process \cite{Masanes2019}. It is therefore required to test Born's rule in experiments, which has been done in terrestrial interferometric experiments, where the bounds for potential deviations have been tightened in increasingly sensitive experiments \cite{PhysRevResearch.3.013296,10.1088/1367-2630/aa5d98}. With a twin experiment, where one interferometer is on the ground and one in space on a satellite, any deviation from Born's rule due to gravitational fields can be verified or ruled out. While such deviation is speculative at the moment, there exist theories about gravitational phase shifts on single photons \cite{Hilweg_2017} and the role of Born's rule in quantum gravity \cite{doi:10.1142/S0217732321500139,Valentini2022}. On the other side, there have been tests of speculative physical theories in the past for which no specific model had existed (such as loophole-free Bell tests \cite{Hensen2015,PhysRevLett.115.250402,PhysRevLett.115.250401,PhysRevLett.119.010402}). In addition, testing Born's rule on the satellite is also a very good tool to test our technology platform, as imperfections of any of our components would seemingly induce deviations from Born's rule.

\section*{Space-qualification}
Each element has been space-qualified on the individual component level, including thermal-vacuum (T-V) cycling, gamma irradiation, as well as mechanical vibration and shock tests. As the launch vehicle is unknown at this time, a Falcon 9 rocket is assumed to estimate the mechanical loads. The vibration tests use different patterns (sine and random) in the frequency bands from 20 to 2000 Hz and levels up to 12 g$_\text{rms}$ for durations ranging from a few seconds up to 8 minutes. The mechanical pyroshock tests were carried out on all three axes sequentially with 60-500 g at frequencies from 0.1 to 10 kHz. For the gamma irradiation we tested for a total ionising dose of 20 krad (typical for LEO). We only tested using ionizing radiation and did not test for displacement damage (which is of more concern for long-term performance). The latter requires particle accelerators which was unfeasible for the amount of our components, as well as their dimensions and shapes. We have also carried out thermal vacuum tests on our payload components to evaluate their operational and non-operational temperature ranges. The complete system will be again qualified using thermal vacuum cycling tests. We would like to mention that here we define the term 'space-qualified' as the component surviving our test procedure while still meeting the minimal operational criteria. This does not necessarily imply that there have been no changes but rather the potential changes are of no concern for the mission's success.

\section*{Payload and system design}
\subsection*{Laser system}
The 698 nm pump laser system to excite the single photon source is adapted from a commercially available ECDL. This wavelength is compatible with hBN emitters having an optical transition above 700 nm \cite{10.1021/acsnano.6b03602,Cholsuk2022}, and it also matches the clock transition of neutral strontium atoms \cite{Poli2014} (therefore it may also prove useful for future space missions with cold atom experiments). A ridge waveguide laser diode is integrated into a butterfly housing with an anti-reflective coated window, together with a thermoelectric cooler (TEC), a thermistor, an external volume Bragg grating at the front facet for controlled optical feedback, and two cylindrical lenses to collimate the output beam. The laser has a free space optical power output of 30 mW at a current of 65 mA. The butterfly package is mounted on an optical breadboard (see Figure \ref{Fig:1}) where the beam is guided through an optical isolator, before being coupled into a polarization-maintaining fiber with an inline power meter used for power stabilization. The laser system has a package size of $45\times 80\times 25$ mm$^3$. We expect a coupling efficiency around 50-60\% into the single mode fiber with this configuration, resulting in around 15 mW of optical power at the fiber output.\\
\indent Our laser driver design is based on the canonical current source with operational amplifier (op-amp) and added laser diode protections. An nMOSFET in series with a current sense resistor and the laser diode controls the current through the laser diode. The voltage drop across the sense resistor is fed to the inverting input of an op-amp with its output driving the nMOSFET. This comprises a stable feedback loop when the output from the digital-to-analog converter (DAC) is connected to the op-amp's non-inverting input. A solid-state relay keeps the potential across the laser diode constant while it is not operating. This prevents electrostatic discharge across the laser diode. A fast-switching diode is connected in parallel to the laser diode to discharge inductive currents when the laser diode is switched off. A capacitor across the laser diode provides a low impedance path for high frequency noise. The driver can deliver up to 150 mA current for a laser diode with 1 V forward voltage and its output current is stable within 5 $\mu$A when tested at 70 mA for a 12 hours period. The complete laser driver electronics fit on a single printed circuit board (PCB) and can be easily mounted into the CubeSat.\\
\indent To qualify the laser against stress, we carried out the mechanical tests outlined above. We characterized the diode lasers by measuring the power-current curve and the optical spectra versus current, before and after the mechanical tests. Supplementary Section S3 shows the results of both tested lasers with negligible changes. The spectral map (see Supplementary Section S3) confirms single frequency operation and a linewidth below 10 pm (limited by the spectrometer resolution). Overall, we report no substantial misalignment of the optical components inside the butterfly package after the mechanical tests. T-V cycling and irradiation are currently ongoing.
\begin{figure*}[t]
\includegraphics[width=0.8\textwidth]{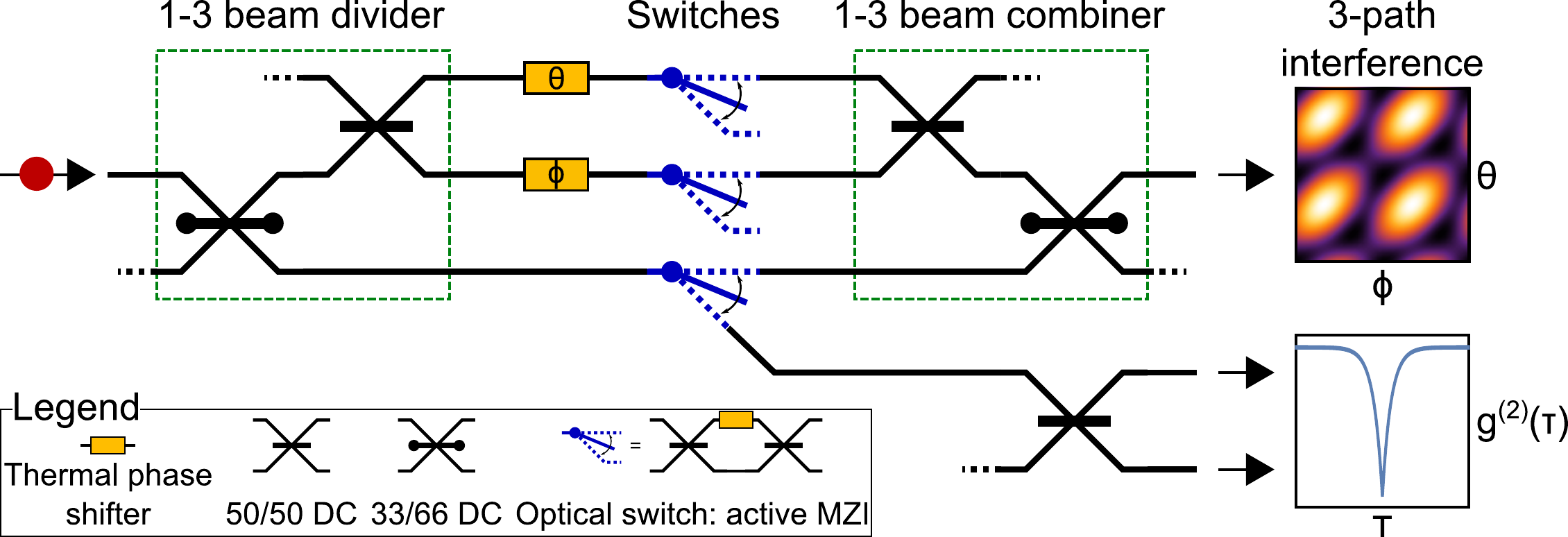}
\caption{Outline of the photonics circuit. A three-arm interferometer is constructed by cascading a 1-to-3 beam divider and a 3-to-1 beam combiner, both composed by directional couplers (DC) with 50/50 and 33/66 splitting ratios. Within the arms of the interferometer, three reconfigurable Mach-Zehnder interferometers (MZIs) act as optical switches that can interrupt the light passage. Two additional phase shifters change the relative optical phases acquired by the light in the different interferometer arms, allowing us to tune the interference pattern. One of the three switches (bottom one) enables us to route the photons to an additional 50/50 DC for measuring $g^{(2)}(\tau)$.}
\label{Fig:2}
\end{figure*}

\subsection*{Quantum photonics module}
The quantum photonics module is based on a room temperature hBN quantum emitter coupled to a PIC fabricated with femtosecond laser micromachining (FLM) \cite{10.1038/nphoton.2008.47}. The quantum emission process in hBN is based on a fluorescent defect that induces an effective two-level system into the bandgap \cite{nnano.2015.242}. When excited non-resonantly with a laser, a single photon is emitted when the system transitions back into the ground state. These emitters can be fabricated by e.g.\ plasma etching \cite{10.1021/acsphotonics.8b00127}, direct laser writing \cite{doi:10.1021/acsphotonics.0c01847}, mechanical damage induced by atomic force microscope tips \cite{doi:10.1021/acs.nanolett.1c02640}, or localized electron \cite{arXiv:2208.13488} and ion irradiation \cite{doi:10.1021/acsomega.1c04564}. The emission process was found to be radiation-tolerant, i.e., hBN quantum emitters can operate in space environments \cite{Vogl2019}. Single photons emitted from hBN have been utilized for use in quantum information processing \cite{White_2021,Conlon2022}, for extended quantum theory tests \cite{PhysRevResearch.3.013296}, and quantum key distribution \cite{https://doi.org/10.1002/qute.202200059}. The integration of hBN emitters with glass fibers where the emitter was placed onto the fiber facet has been demonstrated already \cite{10.1088/1361-6463/aa7839}. This configuration has the advantage of the emitter being directly interfaced with the waveguiding element without the need for any additional optics in between. Moreover, the higher refractive index of the dielectric environment of the collection optics modifies the dipole emission pattern which results in an increased emission into the waveguide \cite{C9NR04269E}.\\
\indent This emitter integration will be applied to glass PICs fabricated by FLM, where ultrashort laser pulses are focused into a transparent dielectric to induce a permanent refractive index increase localized at the laser focus. In this way, waveguiding structures can be written directly along arbitrary three-dimensional paths, without the need for any lithographic masks. Laser written PICs have been employed for quantum light processing \cite{Corrielli2021}, and have also been qualified for use in space environments with respect to their radiation tolerance and their insensitivity to temperature or pressure changes \cite{https://doi.org/10.1002/lpor.202000167}. Using FLM, several integrated optical elements can be produced, including beam splitters (directional couplers) \cite{Corrielli:18}, phase and polarization rotators \cite{Corrielli2014,Lammers:19}, and Mach-Zehnder interferometers (MZIs) \cite{Chaboyer:17}. The latter can be actively tuned by integrated thermal phase shifters that control the relative phase difference between the MZI arms. These devices consist of resistive elements heating specific waveguide segments locally and thereby increasing their effective refractive index via the thermo-optic effect. To avoid crosstalk between neighboring thermal shifters and to increase their efficiency, pairs of deep isolation trenches can be machined by laser ablation around the resistive elements, which confine the dissipated heat at the underneath waveguide \cite{https://doi.org/10.1002/lpor.202000024}. Alternatively, U-shaped undercuts, such that the waveguides with the resistive elements are suspended bars, have a higher isolation. For testing the mechanical stability of these devices, we have subjected test structures with different sizes to pyroshock and vibration tests. The smallest suspended bars with a width of 20 $\mu$m had a failure probability of 6.67\% (see Supplementary Section S4), while suspended bars with a width of 80 $\mu$m have all survived. In addition, all trench pairs have survived as well, independently of their size.  For the QUICK$^3$ circuit, we therefore implement thermal shifter elements with 90 $\Omega$ resistors included between isolation trenches with 3 mm length (along the waveguide direction), 300 $\mu$m depth and 100 $\mu$m width, and 27 $\mu$m separation between both trenches. This configuration allows to produce a 2$\pi$ phase shift in vacuum with $<$10 mW electrical power dissipation, a rise/fall time in the order of 100 ms, and a thermal crosstalk between neighboring shifters $<$5\%. This remaining thermal crosstalk can be fully characterized and taken into account during operation.\\
\indent The schematic circuit layout is depicted in Figure \ref{Fig:2} (for the actual layout see Supplementary Section S5). For testing Born's rule, we need a three-path interferometer where we can block each path individually. We first split the incoming photon beam into three paths with equal amplitude using a 1-to-3 beam divider composed by two directional couplers. In each path we have a MZI switch that can route the photons out of the interferometer, therefore allowing us to selectively block the light passage through that path. The photons interfere at a 3-to-1 beam combiner where a single output mode is guided to a single photon detection module. For each path there are additional phase shifters to tune the interferometer to an interference maximum in the phase diagram (see Supplementary Section S6). To verify the general functionality of the single photon source, we can route the photons through the circuit to a balanced directional coupler, which allows us to measure the second-order correlation function. Both outputs of this directional coupler are also connected to two single photon detectors.\\
\indent By considering state-of-art performances of the waveguides fabricated with FLM, the design footprint of the circuit layout is $45\times 24\times 1$ mm$^3$ with expected insertion losses $<$2 dB for the single mode fibers used for light in- and out-coupling. All directional couplers can be fabricated by FLM with very tight tolerances on the splitting ratio ($<$1\%) and this ensures a high visibility of the 3-path interference pattern as well as high extinction ratio of the optical switches ($>$30 dB).

\subsection*{Laser suppression}
To suppress the excitation laser and avoid saturation of the single photon detectors, we implement a transmission filter based on a fiber Bragg grating (FBG), consisting of a refractive index modulation in the fiber core that connects the quantum photonics module with the detectors. The FBG acts as a narrowband integrated mirror, whilst the reflection wavelength is defined by the period of the grating and the effective refractive index of the guided mode. Additionally to the transmission dip of the reflected resonance wavelength, however, the transmission spectrum also shows resonances on the short wavelength side caused by cladding mode resonances, where light is reflected into modes guided in the cladding of the fiber. This cross coupling can be suppressed by increasing the size of the cross-section of the homogeneous refractive index modulation to be larger than the fiber core \cite{1316949,Thomas:11}. The grating inscription can be done by FLM as well. The periodic modulation is generated using the phase mask technique, where a pure two-beam interference pattern is generated by a specially designed phase mask, which is then imprinted into the fiber with a cylindrical lens \cite{https://doi.org/10.1002/lpor.201100033}. The design parameters for the transmission filter are a resonance wavelength around 690 nm (defined by the ECDL), with a bandwidth around 500 pm and a minimum suppression of -60 dB, while keeping the broadband losses below -1 dB. The FBGs will be inscribed using a phase mask with a pitch of 1420 nm, resulting in an FBG period of 710 nm, with the third-order reflection \cite{https://doi.org/10.1007/s00339-010-6065-6} at 690 nm. To achieve the high suppression and target bandwidth, the grating will have a length of around 40 mm.\\
\indent As already mentioned, the optical elements are connected to the waveguide with single mode fibers, where the fibers serve additional purposes (such as an in-line power meter for monitoring or the above mentioned transmission filter). The idea is to glue the fibers directly to the waveguide in- and outputs. Due to all components being single mode, mode matching is no issue. Our fiber-connected waveguides have been subjected to the mechanical qualification tests as well. While there was one device showing a higher variation (i.e., higher than our measurement uncertainty) in the end-to-end coupling efficiency, all other devices have shown no changes (see Supplementary Section S4). Moreover, also the 'damaged' chip still had a tolerable coupling efficiency. We can therefore conclude that our glued fiber-to-waveguide connections can be applied in QUICK$^3$.

\subsection*{Single photon detectors}
The single photons that passed through the PIC are detected by a single photon detection system with three channels. In recent years, there have been several developments toward satellite-based single photon detectors based on actively quenched single photon avalanche diodes (SPADs) \cite{doi:10.1126/science.aan3211,Ren2017,Yang:19}. Targeting smaller spacecrafts, passively quenched SPADs were operated on-board the CubeSats Galassia and SpooQy-1 to measure the statistics and violation of Bell's inequality of entangled photon pairs \cite{Villar:20,PhysRevApplied.5.054022}.\\
\indent For QUICK$^3$, we build on the heritage of the single photon detection system used in these missions. While the SPADs (SAP500) and the power supply circuit remained largely identical, the detection circuits were upgraded with an active quenching stage that reduces the dead time (from $>$1 $\mu$s to nominal 60 ns) to increase the detection rates above 10$^6$ photons/s in saturation. Further, the improved detection system features a TEC for each detector that suppresses dark counts by cooling the SPADs. The system (designed and built by SpeQtral Pte.\ Ltd.\ in Singapore) consists of three SPADs and a time-interval counter in a compact module of around 0.85U. The detectors together with their active quenching stage are secured in an aluminum housing that acts as a mechanically stable structure and as the heat sink. The optical input is fiber coupled with FC/PC terminated connectors. Power and time tagging circuits are hosted on a dedicated PCB that is connected to the aluminum housing. The unit is controlled by a single-board computer (Xiphos Q7). Time-intervals between two of the detectors can be stored in the integrated storage and retrieved to produce the second-order correlation histogram. Finally, communication with the payload controller is handled via SPI (serial peripheral interface). The detection efficiency is adjustable between 0-50\% by setting the detector bias voltage. The timing resolution of the system is limited by the single photon detection jitter of around 500 ps. The detector dark count rate depends on the thermal environment and on the (for cooling) available power. For an environment near room temperature, we expect dark count rates below 1 kHz per detector prior to launch (depending on the overvoltage setting). The dark count rate will slowly increase over time due to radiation damage in orbit \cite{10.1117/12.2583934}. Due to the flexibility of the operational concept of the module (i.e., adjustable cooling and overvoltage), the power consumption depends on the mode of operation with a peak power consumption of 12.5 W for saturation of all detectors and maximum cooling.

\subsection*{Electronics}
The experiments are operated through a payload controller that implements all interfaces with the different components. This consists of power distribution, serial interfaces toward the laser and detector electronics, and an eight-channel current driver for the optical switches on the PIC. The data from the experiments is collected and forwarded to the satellite's onboard computer via a serial communication interface. Building upon the experience from the TUBiX-20 missions \cite{BARSCHKE2020108,Gordon2020}, the reliability of the controller is enhanced by implementing current limits and an external watchdog timer as protection mechanisms. Following a careful commercial off-the-shelf  approach \cite{Sinclair2013}, the used components are either proven by having flight heritage or total ionizing dose tested.\\
\indent The eight PIC switches are controlled by the resistive heating elements on the waveguides which are actuated by an op-amp-based current driver. The output voltage of an eight-channel DAC is compared to the voltage across a low-side current sense resistor by the op-amp. An n-channel MOSFET that is driven by the output of the op-amp controls the current through the optical switch and thus closes the feedback loop. The setup is capable to drive eight individual resistive loads (in the range of 70-110 $\Omega$) with each up to 23 mA of current at a resolution of 23 $\mu$A. Additionally, the actual current is measured over the driver's current sense amplifier using an analog-to-digital converter which is used for monitoring and calibration.
\begin{figure*}[t]
\includegraphics[width=0.8\textwidth]{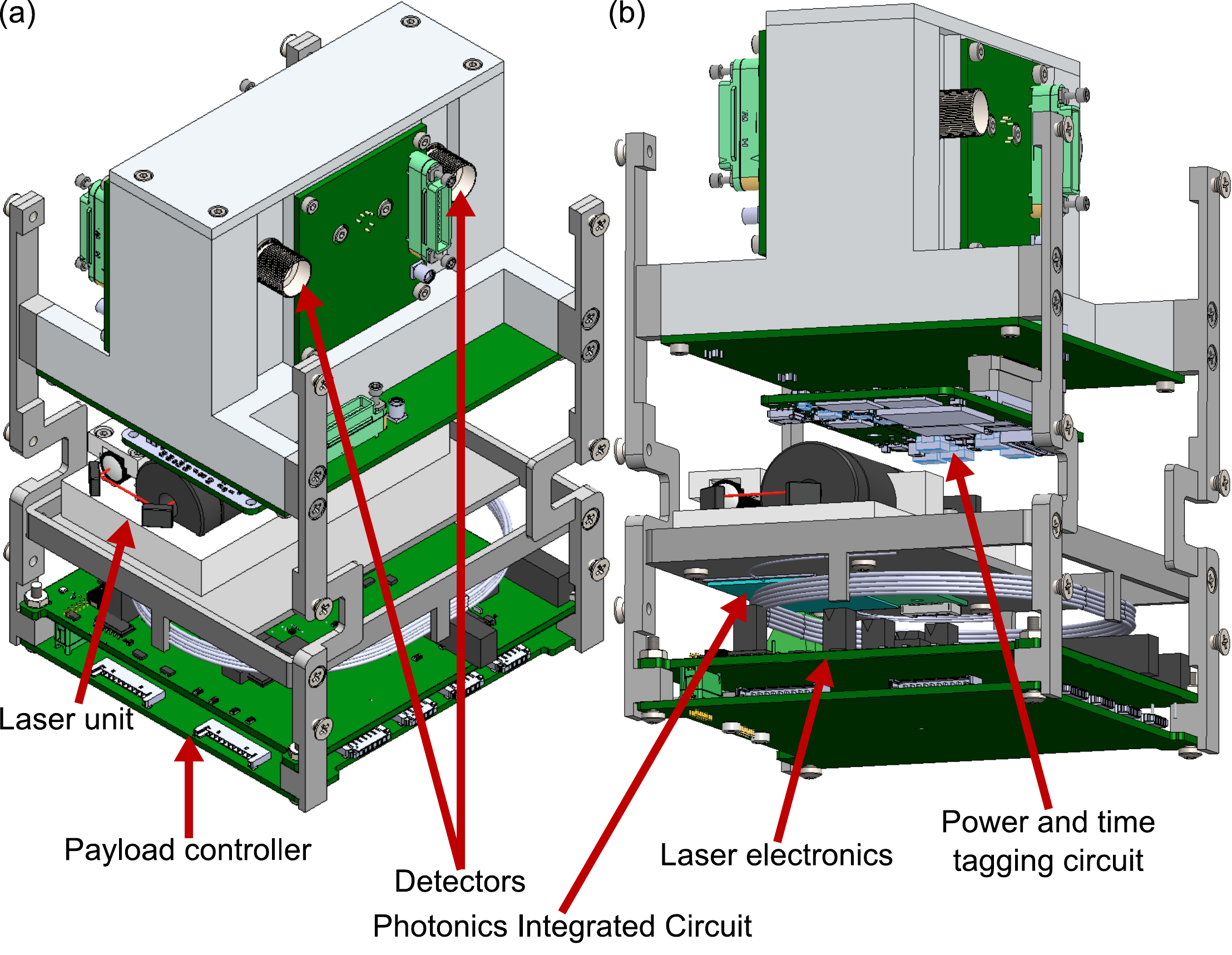}
\caption{(a) Assembly of the payload components in the sub-frame that is going to be mounted in the CubeSat main structure. (b) Assembly from a different perspective showing the photonics integrated circuit.}
\label{Fig:3}
\end{figure*}

\subsection*{Payload assembly}
All payload sub-systems are first mounted into a sub-frame that can then be inserted as a complete package into the CubeSat main structure. This has the advantage that we can pre-assemble all components into our sub-frame, which simplifies the integration with the spacecraft. Figure \ref{Fig:3} depicts the payload assembly: our payload controller (the spacecraft-to-payload interface) is mounted at the bottom of our sub-frame. The laser electronics are comprised of a single PCB and supply power to the laser unit as well as handle the communication between the payload controller and the laser. The laser unit itself is mounted on an additional aluminum plate that also serves as the holder for the PIC on the bottom side and holds the electrical interface to the PIC switches. The detector unit with the power and time tagging circuit below the housing of the detectors is mounted at the top of the payload sub-frame. The detector's central placement is due to the need to route fibers to both sides of the detector unit. For clarity, not all optical fibers and electrical cables running along the sides of the components are shown in Figure \ref{Fig:3}, but were taken into account in the design.

\section*{Coupling with quantum memories}
Crucial for future global quantum networks is the storage of single photon states in quantum memories. Recent works have demonstrated that space-borne quantum memories (QM) would not only enable truly global quantum communications in an untrusted fashion \cite{Gundogan2021,Liorni_2021} but also act as a platform to probe the intersection of quantum physics with gravity \cite{arXiv:2209.02099}. Storing single photons in long-lived QMs would thus constitute an important step towards these important applications.\\
\indent For the coupling of single photons to quantum memories, hBN emitters are a promising quantum light source. We have predicted a rich variety of fluorescent defects in hBN with transition energies compatible with many quantum memory platforms using density functional theory (DFT) \cite{Cholsuk2022}. We decided to work with alkali vapor-based QMs since devices with sufficiently small footprints have recently been developed for space applications \cite{Strangfeld:22}. Furthermore, alkali vapor platforms can offer broad bandwidth operation around 1 GHz \cite{PhysRevLett.116.090501,PhysRevLett.119.060502}. The linewidth of hBN single photons can be reduced to match these systems (even at room temperature) upon integrating them into tunable optical microcavities \cite{10.1021/acsphotonics.9b00314} (we note that we are currently building an improved version of the initial cavity-coupled emitter that will feature an even narrower linewidth). Moreover, the tunability of such microcavities allows us to adapt the resonance wavelength to match exactly the atomic transition. By investigating 267 different emitters using DFT, we have identified the following defects emitting at relevant transitions in alkali vapors \cite{Cholsuk2022}:
\begin{itemize}
\item Na-D$_2$: O$_\text{B}$V$_\text{B}$
\item Na-D$_1$: Ti$_\text{N}$V$_\text{B}$ and V$_\text{N}$V$_\text{B}$Ti
\item Rb-D$_2$: Er$_\text{B}$V$_\text{B}$ and Er$_\text{N}$V$_\text{B}$          
\item Rb-D$_1$: In$_\text{B}$V$_\text{B}$ and Er$_\text{N}$N$_\text{B}$V$_\text{N}$ and Al$_\text{B}$V$_\text{B}$
\item Cs-D$_2$: Er$_\text{N}$V$_\text{N}$
\item Cs-D$_1$: Er$_\text{B}$V$_\text{B}^{+}$
\end{itemize}
Here the defects are denoted in Kr\"oger-Vink notation where e.g.\ the last listed defect complex consists of an erbium impurity replacing a boron atom next to a boron vacancy with a positive charge state. This direct compatibility of microcavity-coupled hBN emitters to alkali-based quantum memories renders the need for frequency conversion (between the light source and the storage medium) unnecessary.

\section*{Conclusion}
We presented the design of a satellite-based quantum light source for quantum communication and extended physical theory tests in space. We note that we might adapt the design based on the current manufacturing and characterizations of the payload. This payload will be integrated into a commercial 3U CubeSat and is part of the QUICK$^3$ mission that is planned for launch in 2024. QUICK$^3$ will demonstrate the first true single photon source in space as a use-case for future QKD networks. In addition, we will also perform a science experiment, where we test a fundamental postulate of quantum mechanics, namely Born's rule, in a space environment. Based on the here presented payload design, we have now begun to manufacture our payload and initiated the first system level tests. The general concept of our quantum optical components for small satellite missions is a promising technology also for other future space missions toward commercial QKD networks. In addition, many building blocks are not unique to QUICK$^3$, e.g., the laser system may be useful for cold atom experiments with strontium and the detector system could be used to realize inter-satellite quantum communication links. As we also explore concepts to couple our satellite-based quantum light source to quantum memories, we expect that the quantum light source will be applied in various scenarios in future quantum networks.

\subsection*{Funding}
This project is funded by the German Space Agency DLR with funds provided by the Federal Ministry for Economic Affairs and Climate Action BMWK under grant numbers 50WM2165, 50WM2166, 50WM2167 (QUICK3), and 50RP2200 (QuVeKS). The authors received funding from the Deutsche Forschungsgemeinschaft (DFG, German Research Foundation) - Projektnummer 445275953. The authors acknowledge support from the Federal Ministry of Education and Research (BMBF) under grant numbers 13N16292 (ATOMIQS) and 13N16028 (MHLASQU). This research is supported by the Research Centres of Excellence programme supported by the National Research Foundation (NRF) Singapore and the Ministry of Education, Singapore. C.C.\ acknowledges a Development and Promotion of Science and Technology Talents Project (DPST) scholarship by the Royal Thai Government. M.G.\ acknowledges funding from the European Union's Horizon 2020 research and innovation programme under the Marie Sk\l{}odowska-Curie grant agreement No.\ 894590.

\subsection*{Availability of data and materials}
Any data and materials, except for details on the commercial spacecraft, are available from the authors upon reasonable request.

\subsection*{Competing interests}
The authors declare that they have no competing interests.

\subsection*{Consent for publication}
All authors have read and agreed to the published version of the manuscript.

\subsection*{Authors' contributions}
T.V. and M.K. have conceived the mission. P.W. and J.B. designed the interface to the satellite bus. L.W. designed the electronics. J.L. designed the mechanical payload assembly. S.Sc., E.D.R., and M.K. designed and qualified the main laser system. N.A., J.K., S.R., M.A., and J.B. qualified backup laser diodes and the quantum photonics module. R.K. and S.N. designed the laser suppression filter. M.G. developed the memory concept. C.C. performed the density functional theory simulations. A.Lo. designed the detector system. S.Sa., D.P., and A.Li. designed the laser driver. G.C. and S.A. designed the photonic integrated circuit. T.V., M.K., P.W., and J.B. supervised the project. All authors contributed to the writing of the manuscript.

\providecommand{\noopsort}[1]{}\providecommand{\singleletter}[1]{#1}%

\end{document}